\documentstyle[psfig,epsf,12pt]{article}
\newcommand{\sect}[1]{\section{#1}\setcounter{equation}{0}}

\def\be{\begin{equation}}
\def\ee{\end{equation}}
\def\ben{\begin{eqnarray}}
\def\een{\end{eqnarray}}
\begin{document}

\begin{titlepage}

\begin{centering}
{\Large {\bf Singularity-free cosmological solutions\\[1mm]
in quadratic gravity\\}}

\vspace{0.4in}
{\bf P. Kanti}, {\bf J. Rizos} and {\bf K. Tamvakis} \\
\vspace{0.3in}
{\it Division of Theoretical Physics,
Physics Department,}\\[1mm] {\it University of Ioannina,
Ioannina GR 451 10, Greece.} \\

\vspace{.4in}
{\bf Abstract} \\
\vspace{.1in}

\end{centering}
{\small
We study a general field theory of a scalar field coupled to gravity through
a quadratic Gauss-Bonnet term $\xi(\phi)\,R^2_{GB}$. The coupling function
has the form $\xi(\phi)=\phi^n$, where $n$ is a positive integer. In the
absence of the Gauss-Bonnet term, the cosmological solutions for an empty
universe and a universe dominated by the energy-momentum tensor of a scalar
field are always characterized by the occurrence of a true cosmological
singularity. By employing analytical and numerical methods, we show that,
in the presence of the quadratic Gauss-Bonnet term, for the dual case of even
$n$, the set of solutions of
the classical equations of motion in a curved FRW background includes
singularity-free cosmological solutions. The singular solutions are shown
to be confined in a part of the phase space of the theory allowing the
non-singular solutions to fill the rest of the space. We conjecture that
the same theory with a general coupling function that satisfies certain
criteria may lead to non-singular cosmological solutions.}\\[6mm]
\vspace*{0.4in}

\vspace{-0.1in}

\begin{flushleft}
\begin{tabular}{l} \\ \hline
{\small Emails~:~pkanti@cc.uoi.gr, irizos@cc.uoi.gr, tamvakis@cc.uoi.gr}
\end{tabular}
\end{flushleft}

\end{titlepage}
\noindent
\centerline{INTRODUCTION}\\

Despite the successes of Einstein's theory of Gravitation at large distances,
a Quantum Theory of Gravity,
valid at supersmall distances, requires a more general framework. At present,
Superstring Theory~\cite{witten} seems to provide the most
appealing framework for such a theory. Superstring Theory leads to the
unification
of gravity with the other fundamental forces. It also leads to important
modifications of the standard cosmology, based
 on the Einstein action, at short distances of the order of the Planck length.
Although the theory has not been fully developed
 to the point that a detailed cosmology could be constructed, a number of
general conclusions can be drawn regarding new
 possibilities that distinguish string cosmology from the
 standard model. Modifications of gravity of stringy origin
 can be studied through the Superstring Effective Action corrected by
incorporating loop and $\alpha'$ effects. The latter are
 associated with the contribution of the infinite tower of massive string
modes, while the former are due to quantum loop effects.
 Although the full String Theory is approximated only in a perturbative sense
by the Effective Action and this is expected to
 describe Physics only up to energies where quantum gravitational effects
start becoming dominant, it is hoped that the loop-corrected Action captures
many of the true features of the exact theory.
The study of the loop-corrected Superstring Action has uncovered
 interesting possibilities~\cite{kaloper} not realized by the
Einstein-Hilbert action such
as the existence of novel stable dilatonic black holes~\cite{kanti1}
 that circumvent the ``no hair" theorem in its restricted sense.
There are of course alternative approaches to
 string cosmology. The absence of cosmological singularities in the presence
of higher-curvature terms, in various numbers of spacetime dimensions, has
been pointed out in the literature several times~\cite{mukhanov}. Another
approach is the pre-Big Bang scenario~\cite{veneziano} which attempts to
incorporate features of the exact theory such as duality symmetries.

 A remarkable property of the loop-corrected Superstring Effective Action
in the presence of the dilaton and moduli fields is
 the existence of {\it{singularity-free}} solutions with flat initial
asymptotics~\cite{anton}. These are linked to the $R^2$ gravitational
 terms with field-dependent coefficients that are present. These solutions
which avoid the initial singularity
 are possible for a definite sign of the corresponding trace anomaly for
which the strong energy conditions related to the
 modulus energy-momentum tensor can be violated. They start from a flat
space-time in the infinite past, they pass
 through an inflationary period and they end up as a slowly expanding universe.
A general field theory of a scalar field
 coupled to gravity through a quadratic Gauss-Bonnet term $\xi (\phi)R^2_{GB}$
has also been shown to possess singularity-free
 solutions in a spatially flat FRW background under very mild assumptions
on the coupling function $\xi(\phi)$~\cite{rizos}. In a subsequent paper by
Easther and Maeda~\cite{maeda}, the case of a closed FRW universe based on
the loop-corrected Superstring Action was also shown to lead, via numerical
methods, to such singularity-free solutions.

In the article at hand, being inspired by the Superstring Effective Theory,
we consider a generic theory with a scalar field
coupled to gravi\-tation through the higher-curvature quadratic Gauss-Bonnet
term. For simplicity, we keep only the one-loop gravitational quantities
that appear in the action functional of the Superstring Effective Theory.
For the case of a spherically symmetric background that we are going to
consider, the other one-loop gravitational quantity $R \tilde{R} \equiv
\eta^{\mu\nu\kappa\lambda} R_{\,\,\,\,\mu\nu}^{\sigma\tau}
R_{\kappa\lambda\sigma\tau}$ vanishes identically. This model captures the essential features and was shown to possess
non-singular solutions in the flat case for a ge\-neral class of coupling
functions~\cite{rizos}. Here, we extend the analysis of Ref.~\cite{rizos}
to the case of a curved universe,
both open or closed. We develop a purely analytical argument and we manage to
show that the singular cosmological solutions, characterized by a true
singularity at finite time, are indeed present in the theory but they are
confined in a certain part of the phase space of the theory. In this way, the
non-singular cosmological solutions are summoned to fill the rest of the space.
  These results are radically different from those that follow from the same
theory when the quadratic Gauss-Bonnet term is absent. In that case, as we
will show, the singular cosmological solutions cover the whole phase space
of the theory leaving no room for the existence of non-singular solutions.

The structure of this article is as follows: In section 1, we derive
the equations of motion for the scalar and gravitational fields in a curved
FRW background. In section 2, we study the cosmological solutions
of the theory when the Gauss-Bonnet term is absent. We consider both the
cases of an empty universe and a universe dominated by the energy-momentum
tensor of a scalar field. In section 3, we develop our analytical argument
for the existence of non-singular cosmological solutions in the presence of
the Gauss-Bonnet term. In section 4, a numerical analysis for a specific
choice of the coupling function serves as an illuminating example for our
theory. The last, short section is devoted to our conclusions.
\vspace*{8mm}
\sect{Equations of motion of the theory}
\noindent

We consider the quadratic coupling of a scalar field with gravity through the
Gauss-Bonnet term which is described by the action
\begin{equation}
S=\int d^4x \sqrt{-g} \left\{ \frac{R}{2} +\frac{1}{2}\,\partial_{\mu}\phi
\,\partial^{\mu} \phi - \frac{1}{16}\,\delta \xi(\phi)\,{\cal R}^2_{GB}
\right\}\,\,.
\label{action}
\end{equation}
Note that $\xi(\phi)$ is, for the time being, a general coupling function.
The Gauss-Bonnet term is defined as
\begin{equation}
{\cal R}^2_{GB}=R_{\mu\nu\rho\sigma} R^{\mu\nu\rho\sigma}-
4 R_{\mu\nu} R^{\mu\nu} + R^2
\end{equation}
and $\delta$, which in Superstring Effective Theory is proportional
to the trace anomaly of the theory, plays the role of a coupling parameter.

The spacetime background assumes the standard spherical symmetric FRW form
\begin{equation}
ds^2=dt^2-e^{2\omega(t)}\,\left\{\frac{dr^2}{1-kr^2}+
r^2\,(d\theta^2+sin^2\theta \, d\varphi^2)\,\right\}\,\,,
\end{equation}
where $k=0, \pm1$ corresponding to the flat, closed and open universe,
respectively.

Making use of the above metric components and assuming further that the scalar
field $\phi$ depends solely on the time coordinate $t$, the equations of
motion take the form
\begin{eqnarray}
&~& \ddot \phi +3\,\dot\phi\,\dot\omega -24\,\frac{df}{d\phi}\,(\ddot
\omega + \dot\omega^2)\,(\dot\omega^2 +k e^{-2\omega})=0\,\,,
\label{eq1}\\[3mm]
&~& 2\,(\ddot\omega + \dot\omega^2)\,(1+8\dot{f} \dot\omega) +
(\dot\omega^2 +k e^{-2\omega})\,(1+8 \ddot{f}) +\frac{1}{2}\,
\dot\phi^2=0\,\,, \label{eq2}\\[3mm]
&~& 3\,(\dot\omega^2 +k e^{-2\omega})\,(1+8 \dot{f} \dot\omega)
-\frac{\dot\phi^2}{2}=0\,\,, \label{eq3}
\end{eqnarray}
where $f(\phi)=-\delta \xi(\phi) /16$. If we set $x=\dot\phi$,
$z=\dot\omega$ and $y=e^{-2 \omega}$, we obtain
\begin{eqnarray}
&~& \dot{x}+3zx +\frac{3}{2}\,\delta \xi'(\dot{z}+z^2)\,
(z^2+k y)=0 \label{equ1}\\[3mm]
&~& (\dot{z} +z^2)\,(4-2\delta\xi'xz)+(z^2+k y)\,(2-
\delta \xi''x^2 -\delta \xi'\dot{x})+x^2=0 \label{equ2}\\[4mm]
&~& (z^2+k y)\,(6-3\delta \xi' x z) -x^2=0 \label{equ3}
\end{eqnarray}
Rearranging eqs.~(\ref{equ1}) and (\ref{equ2}), we obtain a new equation
which contains only the time derivative of $z$
\begin{equation}
\dot{z} = \frac{dz}{d\phi}\,x = -z^2 - \frac{(2-\delta\xi'' x^2 +
3 \delta \xi' z x)\,(z^2+k y)+
x^2}{4-2\delta\xi'zx + \frac{3}{2}\,(\delta\xi')^2 (z^2+k y)^2}
\label{basicz}
\end{equation}
while, from the definition of $y$ we are led to the following differential
equation
\begin{equation}
\dot{y} = \frac{dy}{d\phi}\,x = -2 y z \,\,.
\label{basicy}
\end{equation}
On the other hand, we may solve eq.~(\ref{equ3}) as an algebraic equation
and write $x$, the time derivative of the scalar field $\phi$, as a function
of $z$, $y$ and $\delta \xi'$ in the following way
\begin{equation}
x=-\frac{3}{2}\,\delta \xi' z\,(z^2+k y) +s \sqrt{\biggl[\,\frac{3}{2}\,\delta
\xi'z\,(z^2+k y)\,\biggr]^2 + 6\,(z^2+k y)} \quad , \quad s=\pm 1\,\,.
\label{basicx}
\end{equation}

Note that the set of equations (\ref{basicz})-(\ref{basicy}) is characterized
by an invariance under the simultaneous change of the signs of $z$ and $s$. In
order to clarify this point, we suppose that we have found a solution, for the
choice $s=+1$, described by the set of equations
\begin{eqnarray}
&~&\dot{z} = \frac{dz}{d\phi}\,x_+ = -z^2 - \frac{(2-\delta\xi'' x_+^2 +
3 \delta \xi'z x_+ )\,(z^2+k y)+
x_+^2}{4-2\delta\xi' z x_+ + \frac{3}{2}\,(\delta\xi')^2 (z^2+k y)^2}\,\,,
\label{z1}\\[3mm]
&~&\dot{y} = \frac{dy}{d\phi}\,x_+ = -2 y z \,\,.
\label{y1}
\end{eqnarray}
where $x_+$ stands for the value of $x$ that corresponds to the choice
$s=+1$. Under the transformation $z \rightarrow -z$, the set of equations
(\ref{z1})-(\ref{y1}) is replaced by a new one with $x_-$ in the place of
$x_+$, where $x_-$ corresponds to the choice $s=-1$. This means that if
$x_+$ corresponds to a solution (singular or not) of the equations of
motion, then $x_-$ corresponds to the same solution with the sign of $z$
reversed. For this reason, we may keep fixed the sign of $z$, e.g. $z>0$,
during the analytical treatment of the problem.
\vspace*{8mm}
\sect{The $\delta=0$ case}
\noindent

We first consider the case with $\delta=0$, that is without the Gauss-Bonnet
term. If we further assume that the scalar field takes on a constant value
and set $x \equiv \dot{\phi}=0$ in equations (\ref{eq1})-(\ref{eq3}),
we obtain
\ben
&~&\dot{\omega}^2+k e^{-2\omega}=0\,\,,\\[2mm]
&~&\ddot{\omega}+\dot{\omega}^2=0\,\,.
\een
For $k=0$, we obtain $\omega=const.$ which corresponds to a static universe
with arbitrary radius. On the other hand, for $k=+1$, we must have
$\omega=const.$ and $e^{-\omega}=0$ at the same time, which corresponds
to a static universe with infinite radius. The only interesting case is the
last one, $k=-1$, where we find that $e^{\omega(t)}=a(t) \sim t$. This result
corresponds to a linearly, eternally expanding universe with an initial
singularity,
at $t=0$. Note that the rate of expansion is much larger than in the case
of the ``radiation" ($a \sim t^{1/2}$) or the ``matter" ($a \sim t^{2/3}$)
epoch of the Standard Cosmological Model. This is due to the absence of any
matter content of the universe capable of slowing down the expansion of the
universe.

Next, we allow the scalar field to evolve with time ($\dot{\phi} \neq 0$)
while keeping the
parameter $\delta$ equal to zero. In this case, the set of equations
(\ref{basicz})-(\ref{basicy}) take the form
\ben
\dot{z} &=& \frac{dz}{d\phi}\,x = -3 z^2 -2k y\,\,,\nonumber \\[2mm]
\dot{y} &=& \frac{dy}{d\phi}\,x = -2 y z \,\,,
\label{system1}
\een
where
\be
x=s\,\sqrt{6\,(z^2+k y)} \qquad , \qquad s=\pm 1 \,\,.
\label{xd=0}
\ee

We are going to study separately the cases of flat and curved space~:
\vspace*{5mm}

{\bf A) Flat Space ($k=0$)}~: The solution of
(\ref{system1})-(\ref{xd=0}) with respect to time $t$ takes the form
\ben
&~&\dot{z}=-3z^2 \,\Rightarrow \,z(t)=(c+3t)^{-1}\,\,,\nonumber\\[4mm]
&~&\dot{y}=-2yz \,\Rightarrow \,y(t)=(c+3t)^{-2/3}\,\,,\\[3mm]
&~&\dot{\phi}=s\sqrt{6}z \,\Rightarrow \,\phi(t)=s\,\sqrt{\frac{2}{3}}\,
\,ln(\,c+3\,t\,)+c'\nonumber\,\,
\een
where $c$ and $c'$ are arbitrary constants. The result for the scale
factor of the universe is
\be
e^\omega=a(t)=(c+3\,t)^{1/3}\,\,,
\ee
which corresponds to an expanding universe with a true cosmological
singularity at finite time. Note that, here, the rate of expansion of the
universe is smaller than the corresponding ones during the two epochs of
the Standard Cosmological Model. The sole reason for this result is the
presence of the energy momentum tensor of the free scalar field $\phi$ on
the right-hand side of the Einstein's equations which leads to the slowing
down of the expansion of the universe in a more effective way than the
energy momentum tensor of a perfect fluid.

\vspace*{5mm}

{\bf B) Curved Space ($k=\pm 1$)}~: In this case, the system (\ref{system1})
may be reduced to a single equation~:
\be
\dot{z} = \frac{dz}{dy}\,\dot{y}= \frac{dz}{dy}\,(-2yz)=-3 z^2 -2ky
\,\Rightarrow\, 2y\,dz-3z\,dy=\frac{2ky}{z}\,dy \,\,.
\ee
If we multiply both sides by $z/y^4$, we obtain
\be
d\biggl(\frac{z^2}{y^3}\biggr)=(-k)\,d\biggl(\frac{1}{y^2}\biggr)
\Rightarrow z^2+ky=c_1\,y^3\,\,,
\label{con1}
\ee
where $c_1$ is a positive constant. Substituting the above in the differential
equation of $y$, we get the result
\be
\dot{y}=\frac{dy}{d\phi}\,x=\frac{dy}{d\phi}\,s\,\sqrt{6 c_1 y^3}=
-2y\,\sqrt{y\,(c_1 y^2-k)} \,\Rightarrow
\label{inva}
\nonumber
\ee
\vspace*{2mm}
\be
\Rightarrow \, y=\frac{c_1\,k + c_2^2\,exp\,\biggl\{-2s\,\sqrt{\frac{2}{3}}\,
\phi \biggr\}}{2\,c_1\,c_2\, exp\,\biggl\{-s\,\sqrt{\frac{2}{3}}\,
\phi \biggr\}}\,\,.
\label{resy}
\ee
{}From the above expression as well as from eq.~(\ref{inva}), it is evident
that there is a further invariance of the solutions under the interchange of
the signs of $s$ and $\phi$. As a result, we may keep fixed the sign of $s$,
e.g. $s=+1$, while allowing $\phi$ to take on both positive and negative
values.

A cosmological singularity is encountered when $a(t) \rightarrow 0$ or
equivalently when $ y \rightarrow \infty$. From the expression (\ref{resy}),
we conclude that, when $k=+1$, $y$ goes to infinity for $\phi \rightarrow
\pm \infty$ while, for $k=-1$, a singular behaviour arises only for $\phi
\rightarrow -\infty$. Near the singularities, we may evaluate the approximate
expression of $y$ which can be written in the following way
\ben
&~& {\rm For} \qquad \phi \rightarrow -\infty \,\Rightarrow\, y \simeq
\frac{c_2}{2c_1}\,e^{-\sqrt{\frac{2}{3}}\,\phi}
\qquad ({\rm for}\,\,k=\pm 1)\,\,,\\[4mm]
&~& {\rm For} \qquad \phi \rightarrow +\infty \,\Rightarrow\, y \simeq
\frac{k}{2c_2}\,e^{\sqrt{\frac{2}{3}}\,\phi}
\qquad ({\rm for}\,\,k=+1)\,\,.
\een
The corresponding expressions for $z$ can be easily derived from
eq.~(\ref{con1}) and exist only when the solution for $y$ exists as well.
By making use of the differential equation for $\phi$, eq.~(\ref{xd=0}), we may
deduce the dependence of the scalar field on time $t$ near the singularities
and, consequently, the expression of the scale factor of the universe in the
same region. Then, we obtain
\ben
&~& a(t) \simeq (c'+3\sqrt{c_1}\,t)^{1/3} \qquad ({\rm for}\,\,k=\pm 1)\,\,,
\label{initial}\\[5mm]
&~& a(t) \simeq (c'-3\sqrt{c_1}\,t)^{1/3} \qquad ({\rm for}\,\,k=+1)\,\,.
\label{final}
\een
The above expressions describe also a universe with a true cosmological
singularity
at finite time. We note that for $k=+1$, that is for the case of a closed
universe, there are always two branches of singular solutions with vanishing
$a(t)$. On the other hand, for the options $k=0,-1$ which correspond to the
cases of a flat and open universe, respectively, there is only one branch of
singular solutions. This result is in perfect agreement with the singularity
content of the Standard Cosmological Model. The open and flat universe are
characterized by only one singularity, the initial one, while in the case of
a closed universe we encounter two cosmological singularities, the initial
and the final one.

It is also worth noting that, for the choice $\delta=0$, the group of singular
solutions found above covers the whole phase space of the theory leaving no
space for the existence of non-singular solutions. The final singularity of
the closed universe (\ref{final}) can be avoided only if we choose
$k=0,-1$. On the other hand, the initial singularity (\ref{initial})
disappears only if we set $c_2=0$. Then, we end up with the totally
unrealistic case of a static universe with infinite radius. As a result,
we conclude that, in the absence of the Gauss-Bonnet term, the only
realistic cosmological solutions, that we may obtain in the framework of
the theory (\ref{action}), contain, at least, one true singularity.

\vspace*{8mm}
\sect{The $\delta \neq 0$ case}
\par
In this section, we are going to search for non-singular cosmological
solutions in the pre\-sence of the quadratic Gauss-Bonnet term in the action
functional of the theory. It will be useful for our analysis to search
for violations of the energy conditions~\cite{penrose} that indicate absence
of singularities. Assuming a perfect fluid form for the energy-momentum tensor
of the system, the energy and pressure are defined as~: $T_{00}=\rho,\
T_{ii}=-p\,g_{ii}$. Using the equations of motion (\ref{eq1})-(\ref{eq3}),
the energy conditions take the form
\begin{eqnarray}
 &~& \hspace*{-1.5cm} \rho+p=-2\,(\ddot\omega-k e^{-2\omega})=
2\,(k\,y + {z^2}) \,\,
\frac{B+ 24\,x^2 z^4
    -6\,\delta\xi'' x^4 z^2\,(z^2+k y)}{B}\,\,,\\[3mm]
&~& \hspace*{-1.5cm} \rho+3p=-6\,({\dot\omega}^2+\ddot\omega)=
\frac{36\,x^2\,z^2}{B}\,(8-\delta \xi'' x^2)\,( k y + {z^2})^2\,\,,
\end{eqnarray}
where
\begin{eqnarray}
B&=& {x^4}\,ky + 5\,{x^4}\,{z^2} - 12\,{x^2}\,{k^2}\,{y^2} 
  - 24\,{x^2}\,k y\,{z^2} -12\,x^2\,z^4+ \nonumber \\[3mm]
&~& 108\,k\,y\,z^4 + 36\,{k^3}\,{y^3} + 108\,k^2\,y^2\,z^2 + 36\,z^6
\end{eqnarray}
and where we have used eq.~(1.5) in order to eliminate $\delta \xi'$.
In this form, we may easily prove that for $k=0,+1$ the term $B$, being
a polynomial with respect to $x^2$ with no real roots, is always
positive definite. Thus the energy conditions can only be violated,
leading to non-singular cosmological solutions, for $\delta>0$. For $k=-1$,
the analysis is much more complicated but it can be shown that the energy
conditions are violated for both signs of $\delta$. However, by making use
of numerical as well as analytical arguments, we may show that non-singular
solutions arise only for $\delta>0$, too. As a result, in our analysis,
we may consider $\delta$ to be always positive.

Next, we will try to determine all the singular
solutions of the theory with the singularity occurring at finite time hoping
that they do not cover the whole phase space leaving some room for the
non-singular ones. The whole treatment will be analytical assuming a
polynomial dependence of the coupling function $\xi(\phi)$ on the scalar
field, $\xi(\phi)=\phi^n$ with $n$ being a positive integer greater than
unit. Since a singular solution is characterized by the vanishing of the
scale factor at some finite time,
$a(t)\equiv e^{\omega(t)} \rightarrow 0$, we will always demand that near
the singularity $y\equiv e^{-2\omega} \rightarrow \infty$. In the same region,
the quantity $z\equiv \dot{\omega}$ will be set to approach a constant value,
zero or infinity while the scalar field $\phi$ will be left free to adopt any
possible behaviour.

\par
We are going to concentrate our attention on the study of the following
cases~:\\[8mm]
\noindent
\noindent
\underline{{\bf (I)} \, $z$ = finite $\neq 0$, \,$\phi$ = any, \,$y
\rightarrow \infty$.}\\[5mm]
\indent
If, near the singularity, $z$ remains finite adopting a constant value,
$z(t)=c$, then, from the differential equation for $y$ (\ref{basicy}),
we obtain
\be
\frac{\dot{y}}{y}=-2z \,\Rightarrow\, y=e^{-2\omega} \sim
e^{-2 \int\,z(t)\,dt} \,\Rightarrow\,
R(t) \sim e^{ct}
\ee
which goes to zero only when $ct \rightarrow -\infty$. This means that the
singularity is approached only at infinite time and for this reason it
must be excluded.\\[10mm]
%
%
%
\noindent
\underline{{\bf (II)} \, $z \rightarrow 0$, \,$\phi$ = any, \,$y
\rightarrow \infty$.}\\[5mm]
\indent
In this case, the first derivative of the scalar field with respect to
time takes the form
\be
x=-\frac{3}{2}\,\delta\xi'zky + s\,\sqrt{(\frac{3}{2}\,
\delta\xi'zky)^2+6ky}\,\,.
\ee

We have to consider the following cases:\\[4mm]
\indent
{\bf (A)} \, $(\frac{\textstyle 3}{\textstyle 2}\,\delta\xi'zky)^2
\gg 6ky \,\Rightarrow\, \frac{\textstyle 3}{\textstyle 8}\,
(\delta\xi'z)^2\,ky \gg {\cal O}(1)$. Then, $x$ can be written as
\ben
x&\simeq&-\frac{3}{2}\,\delta\xi' zky +\frac{3s}{2}\,zy \,\delta\,|\xi'
k|+\frac{2sk}{z\,\delta\,|\xi'k|} = \nonumber\\[6mm]
&=& \Biggl\{ \begin{tabular}{l} {\rm if} \quad $ s=+1
\quad \Rightarrow \quad  x_+=\frac{\textstyle 2}{\textstyle
\delta\xi'z}$ \,\,, \\[6mm]  {\rm if} \quad $s=-1
\quad \Rightarrow \quad x_-=-3\delta\xi'zky$ \,\,.
\end{tabular}
\een
The above values of $x$ have been taken for the case $\xi'k>0$.
If we change the sign of $\xi'$ according to $\xi' \rightarrow
-\xi'$, we obtain $x_\pm \rightarrow -x_\mp$. On the other hand, if we
change the sign of $k$ in the same way, we are led to $x_\pm \rightarrow
\mp x_\mp$. Note that, apart from the interchange $x_+ \leftrightarrow x_-$,
the only thing that changes is the absolute sign
of $x$ which is not going to be used in the following analysis. For this
reason, we may consider only the cases $\xi'>0$ and $k>0$.\\

We are going to study each expression of $x$ separately:\\[4mm]
\indent
{\bf (a)} \, $x=x_+=\frac{\textstyle 2}{\textstyle \delta\xi'z}$. Then,
eq.~(\ref{basicz}) reduces to~:
\be
\dot{z}=\frac{dz}{d\phi}\,\frac{2}{\delta\xi'z} \simeq -z^2\,\biggl\{
1-\frac{8\,\delta\xi''}{3\,(\delta\xi'z)^4 ky} \biggr\}\,\,.
\ee
If $\frac{\textstyle 8\,\delta\xi''}{\textstyle 3\,(\delta\xi'z)^4 ky}
\ll {\cal O}(1)$, then, rearranging the differential equations for $z$
and $y$, we obtain~: $2y\dot{z}=z\dot{y}\,\Rightarrow\, z^2 \sim y$
which is inconsistent with our assumption for the behaviour of $z$ near
the singularity. Next, we assume that $\frac{\textstyle 8\,\delta\xi''}
{\textstyle 3\,(\delta\xi'z)^4 ky} =b \simeq {\cal O}(1)$. In the same way,
we obtain $z^2 \simeq y^{(1-b)}$ which is consistent with our assumptions
only for $(1-b)<0$. However, the differential equation for $y$ gives the
result $ y^{(b-1)} \sim (t+c)$, which leads to the conclusion that
the singularity is approached only at infinite time and for this reason it
must be excluded. If, finally, $\frac{\textstyle 8\,\delta\xi''}{\textstyle
3\,(\delta\xi'z)^4\,ky} \gg {\cal O}(1)$, we are led to
\ben
&~&\begin{tabular}{l}
$\dot{z}=\frac{\textstyle 8\,\delta\xi''}{\textstyle 3\,(\delta\xi')^4\,z^2
ky}\, \Rightarrow \,
\frac{\textstyle d\,(z^2)}{\textstyle d\phi}=\frac{\textstyle 8}{\textstyle
3}\,\frac{\textstyle (\delta\xi)''}{\textstyle (\delta\xi')^3 ky}$
\\[7mm] $\dot{y} =-2yz \, \Rightarrow \, \frac{\textstyle d\,(z^2)}
{\textstyle d\phi}=
-\biggl(\frac{\textstyle y'}{\textstyle y\,\delta\xi'} \biggr)^{'}$
\end{tabular} \,\, \Biggr\}
\, \Rightarrow \nonumber \\[6mm]
&\Rightarrow& \, y''\,(\delta\xi')^2 y - y'^2\,(\delta\xi')^2 -
y'\,y\,\delta\xi''\,\delta\xi' + \frac{8}{3k}\,\delta\xi''\,y=0\,\,.
\een
If we assume that $\xi(\phi)=\phi^n$, the only solution of the above
differential equation, compatible with this assumption, is the following
\be
y=\frac{b_1}{\delta\xi''} \quad {\rm and} \quad y'=\frac{b_2}{\delta\xi'}
\ee
provided that
\be
b_1=\frac{8\,(n-1)^2}{3k\,n\,(2-n)} \quad {\rm and} \quad
b_2^2+2\,b_2\,b_1 -\frac{8}{3}\,b_1=0\,\,.
\ee
But, then, we obtain~:
\be
(\delta\xi'z)^2\,y =(\delta\xi')^2\,\Bigl(-\frac{y'}{y\,\delta\xi'}\Bigr)\,
y=-y'\,\delta\xi'=-b_2 ={\cal O}(1)
\ee
which is inconsistent with our assumption that $(\delta\xi'z)^2\,y \gg
{\cal O}(1)$. \\[5mm]
\indent
{\bf (b)} \, $x=x_-=-3\delta\xi'zky$. Then, eq.~(\ref{basicz}) takes
the form \\[1mm]
\be
\dot{z}=\frac{\textstyle dz}{\textstyle d\phi}\,(-3\delta\xi'zky) \simeq
-z^2\,(1-6\,\delta\xi''ky)\,\,.
\ee
If we assume that $6\,\delta\xi''ky \ll {\cal O}(1)$ or $6\,\delta\xi''ky =
b \simeq {\cal O}(1)$, we obtain exactly the same result as in case {\bf (a)}.
The first assumption leads to infinite $z$ while the second one leads to
a singularity which is approached only at infinite time. The third option,
$6\,\delta\xi''ky \gg {\cal O}(1)$, leads to
\be
\dot{z}=\frac{dz}{z\phi}\,(-3\delta\xi'zky) \simeq z^2\,6\,\delta\xi''ky \,
\Rightarrow \, z \simeq (\delta\xi')^{-2}
\ee
which goes to zero only if $\phi \rightarrow \infty$. By using the above
result and for $\xi(\phi)=\phi^n$ with $n>2$, the differential equation for
$y$ gives~: $y \sim (\delta\xi'')^{-1}$ or equivalently that $\delta\xi''y
\simeq {\cal O}(1)$
which is inconsistent with our assumption that $\delta\xi''y \gg {\cal O}(1)$.
For the special case of $n=2$, the same equation leads to $y \sim ln\phi$
and consequently to $(\delta\xi'z)^2 y \ll {\cal O}(1)$ which is again
inconsistent with the assumption $(\delta\xi'z)^2 y \gg {\cal O}(1)$.\\[7mm]
{\bf (B)} \, $(\frac{\textstyle 3}{\textstyle 2}\,\delta\xi'zky)^2
\ll 6ky \,\Rightarrow\, \frac{\textstyle 3}{\textstyle 8}\,
(\delta\xi'z)^2\,ky \ll {\cal O}(1)$. Then, $x \simeq \pm\sqrt{6ky}$, which
means that the following analysis is valid only for $k=+1$. The
differential equation for $z$ takes the form~:\\[1mm]
\be
\dot{z}=\frac{dz}{d\phi}\,x \simeq -z^2 -\frac{(8-6\,\delta\xi''y)\,y}
{4+ \frac{3}{2}\,(\delta\xi')^2 y^2}\,\,.
\ee

Now, we have to consider the following cases concerning the quantity
$\delta\xi'y$ that appears in the denominator of the above equation~:\\[4mm]
{\bf (a)} \, $\delta\xi'y=a={\cal O}(1)$. This means that $\phi \rightarrow
0$ and consequently that $\delta\xi''y \gg {\cal O}(1)$. Then, the
differential equation for $z$ takes the form
\be
\dot{z} \simeq \frac{6\delta\xi''y^2}{A} \, \Rightarrow \,
dz \simeq \pm \frac{\sqrt{6}}{A}\, d(\delta\xi')\,y\,\sqrt{y}\,\,,
\ee
where $A=4+\frac{3}{2} a^2$. Since $\delta\xi'y=const.$, we easily obtain
$d(\delta\xi')\,y=-\delta\xi'\,dy$. Then, we find that $z \sim \sqrt{y}$
which once again leads to a behaviour of $z$ radically different
from the assumed.\\[4mm]
{\bf (b)} \, $\delta\xi'y \gg {\cal O}(1)$. In this case, and according to
our assumption that $\xi(\phi)=\phi^n$, we can only have~: $\delta\xi''y
\gg {\cal O}(1)$. Then, we obtain
\be
\begin{tabular}{l}
$\frac{\textstyle dz}{\textstyle d\phi}\,\,x \simeq \frac{\textstyle
4\,\delta\xi''}{\textstyle (\delta\xi')^2}$\\[7mm]
$\frac{\textstyle d\,(\pm\sqrt{6y})}{\textstyle d\phi}=\frac{\textstyle
dx}{\textstyle d\phi}=-z$ \end{tabular}
\,\,\Biggl\} \,\,\Rightarrow \,\, x'' \, x +
\frac{4\,\delta\xi''}{(\delta\xi')^2}=0 \,\,.
\ee
The only solution of the above differential equation, compatible with our
assumption for $\xi(\phi)$, is the following
\be
x=\frac{b_1}{\sqrt{\delta\xi''}} \quad , \quad {\rm where} \quad
b_1^2=\frac{16\,(n-1)^2}{n\,(2-n)}\,\,.
\ee
However, the above result leads to $\delta\xi''y \sim {\cal O}(1)$ which is
inconsistent with our assumption that $\delta\xi''y \gg {\cal O}(1)$.
Moreover, the above solution for $x$ is real only if $n<2$ which is in
disagreement with our assumption for $\xi(\phi)$.\\[4mm]
{\bf (c)} \, $\delta\xi'y \ll {\cal O}(1)$. In this case, the differential
equation for $z$ takes the simple form $\dot{z} \simeq -(2-\frac{3}{2}\,
\delta\xi''y)\,y$. For the assumptions $\delta\xi''y \ll {\cal O}(1)$ and
$\delta\xi'y=b \simeq {\cal O}(1)$, we obtain the result $z^2 \sim y$
which is different from our assumption that $z$ goes to zero near the
singularity. The other option, $\delta\xi''y \gg {\cal O}(1)$, is a little
more complicated as it leads to
\be
\begin{tabular}{l}
$\frac{\textstyle dz}{\textstyle d\phi}\,\,x \simeq \frac{\textstyle
3}{\textstyle 2}\,\delta\xi'' y^2$\\[7mm]
$\frac{\textstyle d\,(\pm\sqrt{6y})}{\textstyle d\phi}=\frac{\textstyle
dx}{\textstyle d\phi}=-z$ \end{tabular}
\,\,\Biggl\} \,\,\Rightarrow \,\, x'' + \frac{\delta\xi'' x^3}{24}=0 \,\,.
\ee
In the same way, the only solution of the above differential equation, for
$\xi(\phi)=\phi^n$, is the following
\be
x=\frac{b_2}{\sqrt{\delta\xi}} \quad , \quad {\rm where} \quad
b_2^2=\frac{6\,(n+2)}{(1-n)}\,\,.
\ee
This means that the above solution is real only for $n<1$ which is
inconsistent with our assumption for $\xi(\phi)$.\\[7mm]
%
{\bf C)} \, $(\frac{\textstyle 3}{\textstyle 2}\,\delta\xi'zky)^2
\simeq 6ky \,\Rightarrow\, \frac{\textstyle 3}{\textstyle 8}\,
(\delta\xi'z)^2\,ky =\frac{\textstyle 1}{\textstyle a}\simeq {\cal O}(1)$.
Then, $x \simeq \frac{\textstyle 3 \lambda}{\textstyle 2}\,\delta\xi'zky$,
where $\lambda=-1 \pm \sqrt{1+a}$, and eq.~(\ref{basicz}) takes the form~:
\be
\dot{z}=-z^2 \Bigl\{\,(1+2a) -\frac{3\lambda^2}{2}\,\delta\xi''ky\,
\Bigr\}\,\,.
\ee
Since $(\delta\xi'z)^2 y \simeq {\cal O}(1)$, we will always have
$\delta\xi''y \gg {\cal O}(1)$ independently of the behaviour of the
scalar field $\phi$ near the singularity. By using the above in the
differential equation for $z$, we obtain that $z \sim (\delta\xi')^\lambda$
which goes to zero for $(\lambda>0, \phi \rightarrow 0)$ or $(\lambda<0,
\phi \rightarrow \infty)$. The solution of the differential equation for $y$
is $y \sim (\delta\xi'')^{-1} + c$ which leads to $y \simeq {\cal O}(1)$ if
$\phi \rightarrow \infty$ or to $y\,\delta\xi'' \simeq
{\cal O}(1)$ if $\phi \rightarrow 0$. The first of these results is
inconsistent with the assumed behaviour of $y$ near the singularity while
the second one disagrees with our assumption that $y\,\delta\xi''
\gg {\cal O}(1)$.
\par
{}From the study of the first two cases, {\bf (I)} and {\bf (II)}, we
conclude that in a singular cosmological solution the first derivative
of the quantity $\omega$ cannot remain finite or vanish near the singularity.
The only option left is the adoption of an infinite value, just like $y$,
which we are going to study now.\\[8mm]
\indent
\underline{{\bf (III)} \, $z \rightarrow \infty$, \,$\phi$ = any, \,$y
\rightarrow \infty$.}\\[7mm]
\indent
We are going to study the following cases~:\\[4mm]
%
\indent
{\bf (A)} \, $(\frac{\textstyle 3}{\textstyle 2}\,\delta\xi'z)^2\,(z^2+ky)^2
\gg 6\,(z^2+ky) \,\Rightarrow\, (\delta\xi'z)^2 \,(z^2+ky) \gg {\cal O}(1)$.
\\[4mm]
\indent
As in case {\bf (II)}, the change of sign of $\xi'$ or
$z^2+ky$ leads to the interchange $x_+ \leftrightarrow x_-$ as well as
to the change of the absolute sign of $x$. However, the sign of $x$
does not appear anywhere in our analysis while the interchange between the
two expressions of $x$ does not affect our arguments since both of them are
being studied. As a result, we are going to assume again that both of
$\xi'$ and $z^2+ky$ always take on
positive values. Then, under the above assumption, the quantity $x$ assumes
the expressions $x_+=2/(\delta\xi' z)$ and $x_-=-3\delta\xi z\,(z^2+ky)$ for
$s=\pm1$, respectively.\\[4mm]
\indent
{\bf (a)} \, $x=x_+=\frac{\textstyle 2}{\textstyle \delta\xi' z}$. Then,
\be
\dot{z} =\frac{dz}{d\phi}\,\frac{2}{\delta\xi'z} \simeq -z^2\,\Biggl\{1
-\frac{8}{3}\, \frac{\delta \xi''}{(\delta\xi'z)^4\,(z^2+k y)}\Biggr\}\,\,.
\ee
First, we assume that $\frac{\textstyle \delta\xi''}{\textstyle
(\delta\xi'z)^4\,(z^2+ky)} \ll {\cal O}(1)$. Rearranging the system of
differential equations for $z$ and $y$, we easily obtain that $z^2 \sim y$
which means that the two quantities have exactly the same behaviour near
the singularity and, if necessary, their sum may be written as
$z^2 +ky =a z^2$, where $a$ an arbitrary constant.
The differential equation for $z$ gives the result $z^2=(\delta\xi +c)^{-1}$
which goes to infinity when $(\delta\xi +c) \rightarrow 0$. However, we have
to check that this limit takes place at finite time. From the expression
of $x$, we obtain
\be
\frac{d\phi}{dt}=\frac{2}{\delta\xi'}\,\sqrt{\delta\xi+c} \,
\Rightarrow\, (\delta\xi +c) \sim (t+c')^2
\ee
which indeed goes to zero at finite time. It is easy to check further that
our basic assumptions $(\delta\xi'z)^2 \,(z^2+ky) \gg {\cal O}(1)$ and
$\frac{\textstyle \delta\xi''}{\textstyle (\delta\xi'z)^4\,(z^2+ky)}
\ll {\cal O}(1)$ are satisfied provided that $\delta\xi' \neq 0$. As a
result, we conclude that the solution
\be
z^2 \sim y \sim \frac{1}{\delta\xi(\phi)-\delta\xi(\phi_s)}
\ee
is an acceptable singular cosmological solution with the singularity being
approached at finite time.

Next, we consider the case $\frac{\textstyle \delta\xi''}{\textstyle
(\delta\xi'z)^4\,(z^2+ky)}=b\simeq {\cal O}(1)$. Since both $z$ and $y$
are infinite near the singularity and
in order to satisfy the above constraint, we must have $\delta\xi'
\rightarrow 0$. The corresponding solution for $z$ is $z^{-2}=(1-b)\,
(\delta\xi+c)$. If $z^2 \ll y$, we obtain $(\delta\xi'z)^2\,(z^2+ky) \simeq
(\delta\xi'z)^2\,y \gg {\cal O}(1)$ and that $z^2 \sim y^{(1-b)}$. Then~:
\be
\frac{\delta\xi''}{(\delta\xi'z)^4\,y}=\frac{3b}{8} \,\Rightarrow \,
(\delta\xi')^{-2} \sim z^2\,y + c'\,\,.
\ee
The above result leads to $(\delta\xi'z)^2\,y \simeq {\cal O}(1)$ which is
inconsistent with our assumption. Following an exactly similar analysis,
we may show that when $z^2 \gg y$ or $z^2 \sim y$, the above integration
yields a constraint which is again in obvious disagreement with our
assumption that $(\delta\xi'z)^2\,(z^2+ky) \gg {\cal O}(1)$.

Finally, we assume that $\frac{\textstyle \delta\xi''}{\textstyle
(\delta\xi'z)^4\,(z^2+ky)} \gg {\cal O}(1)$. If, furthermore, $z^2 \sim y$,
we may set $z^2+ky=a z^2$ and the differential equation for $z$ gives $z^4
\sim (\delta\xi')^{-2}$. But, this leads to $(\delta\xi'z)^2\,(z^2+ky) \simeq
{\cal O}(1)$ which is inconsistent with our assumption. If $z^2 \gg y$, we
reach the same result as in the case $z^2 \sim y$. Finally, the case $z^2
\ll y$ has been studied in {\bf (IIAa)} and has been shown to lead to the
result $(\delta\xi'z)^2\,(z^2+ky)\simeq {\cal O}(1)$ which is
inconsistent with our assumption.\\[4mm]
{\bf (b)} \, $x=x_-=-3\delta\xi'z\,(z^2+ky)$. In this case, the equation
for $z$ takes the form
\be
\dot{z} \simeq -z^2 \Biggl\{ 1-\frac{6\delta\xi''\,(z^2+ky)^2}
{4z^2 +(z^2+ky)} \Biggr\}\,\,.
\ee

\noindent
We start by assuming that $\frac{\textstyle 6\delta\xi''\,(z^2+ky)^2}
{\textstyle 4z^2 + (z^2+ky)} \ll {\cal O}(1)$. Then, from the differential
equations for $z$ and $y$, we obtain $z^2 \sim y$ which leads to
$\delta\xi''\,z^2 \ll {\cal O}(1)$. Setting $z^2+ky=a\,z^2$ and integrating
the equation for $z$, we find that
$\delta\xi'\,z^2 \sim \phi +...$ from which we conclude that the scalar
field must go to infinity near the singularity. Unfortunately, this result
leads to $\delta\xi''\,z^2 \sim \delta\xi''\,\phi\,(\delta\xi')^{-1}
\simeq {\cal O}(1)$ which is inconsistent with $\delta\xi''\,z^2 \ll
{\cal O}(1)$.

If we assume that $\frac{\textstyle 6\delta\xi''\,(z^2+ky)^2}{\textstyle
4z^2 + (z^2+ky)}=b \simeq {\cal O}(1)$ and, furthermore, that $z^2 \sim y$,
the differential equation for $z$ gives the result~: $\delta\xi'z^2 \sim
\phi+ ...$ which
goes to infinity only when $\phi \rightarrow \infty$. Substituting this
in the same equation, we obtain that $z \sim \phi^{\frac{\textstyle 2a\,
(1-b)}{\textstyle (4a+4b-3ab)}}$. In order to assure the assumed behaviour
of $z$ near the singularity, we demand the positivity of the exponent of
$\phi$. But, then we conclude that $\xi'(\phi)<\phi$ which is inconsistent with
our assumption that $\xi(\phi) =\phi^n$ with $n>1$. For the case $z^2 \gg y$,
we obtain the same result with $a=1$. Finally, if $z^2 \ll y$, the two
constraints become $\delta\xi'y \gg {\cal O}(1)$ and $6\delta\xi''y=b$.
{}From the differential equation for $y$, we obtain that $y\,\delta\xi'
\sim \phi+ ...$ which again goes to infinity only if $\phi \rightarrow
\infty$. Substituting this result in the same equation, we find that
$y \sim \phi^{4/(4+b)}$ with $4+b>0$. But, this leads once again to
the inconsistent result $\xi'(\phi)<\phi$.

Finally, we consider the case where $\frac{\textstyle 6\delta\xi''\,
(z^2+ky)^2}{\textstyle 4z^2 + (z^2+ky)} \gg {\cal O}(1)$. If $z^2 \sim y$,
the differential equation for
$z$ gives $z \sim (\delta\xi')^{-2a/(4+a)}$ which leads to $\delta\xi'z^2
\sim (\delta\xi')^{(4-3a)/(4+a)}$. If we set, $y=(a-1)\,z^2$, the
differential equation for $y$ gives the result $z \sim \phi^{2a/(3a-4)}$.
If $a>4/3$ and $\phi \rightarrow 0$, the constraint $\delta\xi'z^2 \gg
{\cal O}(1)$ is satisfied but $z \rightarrow 0$. The same holds if $a<4/3$
and $\phi \rightarrow \infty$. In both cases, the behaviour of $z$ is
inconsistent with the one that has been assumed. If $z^2 \gg y$, the equation
for $z$ gives
$z \sim (\delta\xi')^{-2/5}$ which goes to infinity only when $\delta\xi'
\rightarrow 0$. But then, $\delta\xi'z^2 \sim (\delta\xi')^{1/5} \rightarrow
0$ which does not agree with our assumption. Finally, if $z^2 \ll y$, we
obtain that $z \sim (\delta\xi')^{-2}$ which goes to infinity only if
$\phi \rightarrow 0$. Then, the differential equation for $y$ gives
$y \sim (\delta\xi'')^{-1} + c$ or equivalently $y\delta\xi'' \simeq
{\cal O}(1)$ which is inconsistent with our assumption. \\[6mm]
%
\indent
{\bf (B)} \, $(\frac{\textstyle 3}{\textstyle 2}\,\delta\xi'z)^2\,(z^2+ky)^2
\gg 6\,(z^2+ky) \,\Rightarrow\, (\delta\xi'z)^2 \,(z^2+ky) \ll {\cal O}(1)$.
\\[4mm]
\indent
In this case, the expression of $x$ becomes $x=\pm \sqrt{6\,(z^2+ky)}$ while
the differential equation for $z$ takes the form
\be
\dot{z} \simeq -z^2\,\Biggl\{1 +\frac{[\,8-6\delta \xi''\,(z^2+ky)\,]\,
(z^2+ky)\,}{4z^2+\frac{3}{2}\, (\delta\xi'z)^2\,(z^2+k y)^2}\Biggr\}\,\,.
\ee

In order to simplify the analysis, we are going to consider first the
possible relation between $z^2$ and $y$ and then study each case
separately.\\[4mm]
\indent
{\bf (a)} \, $z^2 \sim y$. Then, the constraint becomes $\delta\xi'z^2 \ll
{\cal O}(1)$ and the differential equation for $z$ reduces to
\be
\dot{z} \simeq -z^2\,\Biggl[\,(1+2a) -\frac{3}{2}\,a^2\,
\delta \xi''\,z^2 \Biggr]\,\,,
\ee
where we have set $z^2 +ky=a z^2$. If we assume that $\delta\xi'' \ll
{\cal O}(1)$, the above equation gives the result $z \sim e^{\pm (1+2a)\,
\phi/\sqrt{6a}}$ which goes to infinity only if $\phi \rightarrow \pm\infty$.
But, then, $\delta\xi' \rightarrow \infty$ and there is no way that the
constraint $\delta\xi'z^2 \ll {\cal O}(1)$ can be satisfied. A similar result
arises when $\frac{3}{2}\,a^2\,\delta\xi''z^2=b\simeq {\cal O}(1)$. If we
assume that $\delta\xi''z^2 \gg {\cal O}(1)$, we find that $z^2 \sim
(\delta\xi'+c)^{-1}$ which goes to infinity only when $\delta\xi' + c
\rightarrow 0$. But, then, the quantity $\delta\xi'z^2$ goes to infinity
as well and this is inconsistent with our initial assumption.\\[4mm]
\indent
{\bf (b)} \, $z^2 \gg y$. In this case, the results are similar to the ones
of the previous case and they easily arise from them if we set $a=1$.\\[4mm]
\indent
{\bf (c)} \, $z^2 \ll y$. In this case, the constraint becomes
$(\delta\xi'z)^2y \ll {\cal O}(1)$ and the differential equation for $z$
takes the form
\be
\dot{z} \simeq -z^2-\frac{(8-6\delta \xi''ky)\,ky}
{4+\frac{3}{2}\,(\delta\xi'ky)^2}\,\,.
\ee

The above equation has been arisen also and studied in case {\bf (IIB)}.
There, it was shown that, for the cases $\delta\xi'y \gg$ or $\ll
{\cal O}(1)$, there was no real solution of the corresponding differential
equation for $x$ compatible with the assumption that $\xi(\phi)=\phi^n$
with $n \geq 2$. On the other hand, for the case $\delta\xi'y =a \simeq
{\cal O}(1)$, we were led to $z \sim \sqrt{y}$ which is again inconsistent
with our assumption that $z^2 \ll y$.\\[7mm]
%
\indent
{\bf (C)} \, $(\frac{\textstyle 3}{\textstyle 2}\,\delta\xi'z)^2\,(z^2+ky)^2
\simeq 6\,(z^2+ky) \,\Rightarrow\, \frac{\textstyle 3}{\textstyle 8}
\,(\delta\xi'z)^2 \,(z^2+ky)
=\frac{\textstyle 1}{\textstyle a} \simeq {\cal O}(1)$.
\\[4mm]
\indent
Since both of $z$ and $y$ approach infinity near the singularity, the only way
to fulfill the above condition is to have $\phi \rightarrow 0$. The
expression of $x$ becomes $x=\frac{3}{2}\,\lambda\delta\xi'z\,(z^2+ky)$,
where $\lambda=-1\pm\sqrt{1+a}$, while the differential equation for $z$
takes the form
\be
\dot{z} \simeq -z^2 -\frac{(8a -\delta\xi''ax^2)\,(z^2+ky)}{4a-8\lambda
+4(z^2+ky)/z^2}\,\,.
\ee

We consider the cases~:\\[3mm]
\indent
{\bf (a)} \, $z^2 \sim y$ . Then, we may set $z^2+ky=\beta z^2=
\tilde{\beta} y$ and the equation for $z$ reduces to
\be
\dot{z} \simeq -z^2\,\Biggl\{\,1+\frac{4a\beta}{2a-4\lambda+2\beta}
-\frac{3 \lambda^2 \beta^2 \delta\xi''z^2}{2a-4\lambda +2\beta}\,
\Biggr\}\,\,.
\ee

We assume further that $\delta\xi''z^2 \ll {\cal O}(1)$ or equivalently that
$\delta\xi''y \ll {\cal O}(1)$. Then, the differential equation for $y$
gives the result $\delta\xi'y \sim (\phi^{-1}+c)$ which leads to
$\delta\xi'y \simeq {\cal O}(1)$ or, since $y \sim z^2$, to $\delta\xi'z^2
\simeq {\cal O}(1)$. Substituting this in the above equation for $z$, we obtain
that $z \sim e^{-B\phi}$, where $B$ is a combination of $a$, $\lambda$ and
$\beta$. This means that near the singularity, $z \rightarrow 0$ which is
inconsistent with the assumed behaviour of $z$. If we assume that
$\delta\xi''z^2 =b \simeq {\cal O}(1)$, we are led to exactly the same result.
Finally, we consider the case $\delta\xi''z^2 \gg {\cal O}(1)$. The equation
for $z$ leads to the result $z \sim (\delta\xi')^{2\lambda\beta/
(2a-4\lambda+2\beta)}$ and in order to fulfill the condition that
$\delta\xi'z^2 \simeq {\cal O}(1)$ we demand that $2a-4\lambda+2\beta
+4\lambda\beta=0$. By using the dependence of $\lambda$ on $a$, we find
that the solution of the above algebraic equation is~: $\beta=1=-a$. This
means that $z^2 +ky=\beta z^2=z^2$ or equivalently that $z^2 \gg y$ which
is inconsistent with our assumption.\\[3mm]
\indent
{\bf (b)} \, $z^2 \gg y$ . Now, the constraint becomes $(\delta\xi'z^2)^2=
\frac{\textstyle 8}{\textstyle 3a}$ while the differential equation for
$z$ is the same as in the previous case where we have set $\beta=1$. The
first two options, namely $\delta\xi''z^2 \ll {\cal O}(1)$ and
$\delta\xi''z^2 =b \simeq {\cal O}(1)$, lead to the inconsistent result
$z \rightarrow 0$. The third option, $\delta\xi''z^2 \gg {\cal O}(1)$,
leads, as above, to an algebraic equation with the sole solution $a=-1$.
However, from our constraint, it follows that $a$ is always a positive
constant which reveals the inconsistency of this solution as well.\\[3mm]
\indent
{\bf (c)} \, $z^2 \ll y$ . Then, the differential equation for $z$
takes the form
\be
\dot{z} \simeq -z^2\,\Biggl\{\,(1+2a) -\frac{3\lambda^2}{2}\,
 \delta\xi''y\,\Biggr\}\,\,.
\ee
We start by assuming that $\delta\xi''y \ll {\cal O}(1)$. In this case,
the differential equation for $y$ gives the result $\delta\xi'y \sim
\phi+c \simeq {\cal O}(1)$ since, as we concluded above, $\phi
\rightarrow 0$ near the singularity. Then, we obtain that $(\delta\xi'z)^2
y \sim (z^2/y) \ll {\cal O}(1)$ which is clearly in disagreement with
our initial assumption. The same result arises for the case $\delta\xi''y
=b \simeq {\cal O}(1)$. Finally, we assume that $\delta\xi''y \gg
{\cal O}(1)$. Then, as in case {\bf (IIC)}, we obtain that
$y \sim (\delta\xi'')^{-1} + c$ or equivalently that $\delta\xi'' y \simeq
{\cal O}(1)$ which is again inconsistent with our assumption that
$\delta\xi''y \gg {\cal O}(1)$.
\par
So, the only singular solution found, with the singularity occurring at finite
time, has the form
\be
z^2 \sim y \sim \frac{1}{\delta\xi(\phi)-\delta\xi(\phi_s)}
\sim \frac{1}{(t-t_s)^2}\,\,,
\label{singular}
\ee
where $\phi_s$ and $t_s$ stand for the value of the scalar field and time,
respectively, at the singularity. Since $y \equiv e^{-2\omega}$, the scale
factor of the universe in this case behaves near the singularity as
\be
a(t) \sim (t-t_s) \longrightarrow 0 \quad {\rm when} \quad
t \rightarrow t_s
\ee
which corresponds to a linearly expanding universe with a true cosmological
singularity at $t=t_s$. During our analysis, no restriction on the value of
the parameter $k$ has been arisen and, for this reason, the above singularity
can be considered either as an initial or as a final one. Note, that the
rate of expansion is much larger that the corresponding ones during the
``radiation" or the ``matter" epoch of the Standard Cosmological model. The
same result was found in the first section in the case of an empty
universe. However, when the presence of the energy momentum tensor of the
scalar field was taken into account the linear dependence of the scale
factor on time $t$ changed into the milder dependence $\sim t^{1/3}$. Now,
after the addition of the Gauss-Bonnet term in the theory, we find again a
scale factor linearly dependent on time. This could be interpreted in the
following way~: although the addition of the Gauss-Bonnet term has made the
search for singular cosmological solutions much more complicated, the
net result of its presence in the theory is to cancel exactly the
contribution of the energy momentum tensor of the scalar field leaving
behind an ``empty", linearly expanding universe. The only difference
between these two cases is that while the truly empty universe gives rise
only to an open universe, as one would expect, the virtually ``empty"
Gauss-Bonnet universe can be interpreted as a closed or open universe.

The question which arises next is whether the presence of the Gauss-Bonnet
term affects the singular cosmological solutions in such a way that they
are restricted in only a region of the phase space of the theory leaving
some room for the non-singular ones. The singular solutions (\ref{singular})
were determined under the assumptions $x=\frac{\textstyle 2}{\textstyle
\delta\xi'z}$ and $\delta\xi' \neq 0$. Then, $x$ tends to zero only
asymptotically, when the singularity is approached, keeping a definite sign
otherwise. If we assume that the singular solutions are characterized by
$x > 0$, then we must always have $\xi' > 0$ since $\delta$ is always
positive. If $\xi(\phi)=\phi^n$ with $n$ even, the first derivative $\xi'$,
and hence $x$, is positive only if $\phi$ is positive as well. This leads
to the restriction of the singular solutions in the positive $\phi$
half-plane in a $z$--$\phi$ graph leaving the other half for the non-singular
ones. The above argument leads to the selection of even values of $n$ since
for odd values the first derivative of the coupling function retains always
a positive sign, for all values of $\phi$, and this leads to the
non-confinement of the singular solutions in a part of the phase
space of the theory.

Finally, it is worth noting that for the cases $k=0$ and $k=-1$ the singular
solutions are restricted not only in the positive $\phi$ half-plane for $x>0$
and $\delta>0$ but, moreover, in the positive $z$ quarter-plane. This can be
justified if we examine eq.~(\ref{equ3}) written in the form
\be
(z^2+ky)\,(6-3\delta\xi'zx)=x^2 \,\,.
\label{trans}
\ee
If we assume that $k=+1$, then, the value $z=0$ is indeed acceptable by the
above equation which means that, in this case, the singular solutions can
 traverse the axis of $\phi$ at the point where $x^2=6y$ and extend at both
the upper and lower $z$ quarter-planes. On the other hand, if $k=-1$, the
value $z=0$ must be excluded in order to ensure the reality of $x$. This means
that, now, the  singular solutions remain confined in one of the two $z$
quarter-planes according to the sign of $z$ that has been initially chosen.
Finally, if $k=0$, setting $z=0$ in the above equation leads to the result
that the singular solutions traverse the $\phi$-axis at the point where
$x=0$. However, this point is only reached near the singularity where
$z \rightarrow \infty$. As a result, the value $z=0$ is not an acceptable one
for the option $k=0$ and the singular solutions are again restricted in one of
the two $z$ quarter-planes.

\vspace*{8mm}

\sect{Numerical Analysis}
\noindent

In this section, we are going to demonstrate numerically the existence
of non-singular solutions in the presence of the Gauss-Bonnet term in the
action functional of the theory. The analytical study of the previous
section has led to the result that the singular solutions, characterized
by the occurring of a cosmological singularity at finite time, exist in the
context of the theory but they do not cover the whole phase space. As we
will see, it is the non-singular solutions of the theory that are summoned
to fill the rest of the space.

As we have already mentioned, the coupling function of the scalar field
to the quadratic Gauss-Bonnet term is of the form $\xi(\phi)=\phi^n$ with
$n$ even. As an illuminating example, we are going to assume the simpler
case, that is $n=2$, in our numerical analysis. The system of the differential
equations (\ref{basicz})-(\ref{basicy}) together with the algebraic equation
(\ref{basicx}) is numerically integrated yielding a solution for
$z \equiv \dot{\omega}$ and $a=e^\omega$. Since the case $k=0$ has already
been studied in Ref.~\cite{rizos}, we are going to concentrate our attention
on the other two cases, namely $k=\pm1$.

The solution for the case $k=+1$ is depicted in Figures 1 and 2 for a family
of singular and non-singular solutions. As we have analytically proven in the
previous section, the singular solutions, in a $z$-$\phi$ graph, are confined
in the positive $\phi$ half-plane, with the $z$-axis playing the role of a
barricade for them, since we have chosen $\delta=0.5>0$. Changing gradually
the boundary conditions of the numerical integration, the singular solutions
cease to exist as we approach the $z$-axis while the non-singular solutions
start to develop and cover the negative $\phi$ half-plane. According to our
analysis of section 3, the singular solutions are characterized by the
simultaneous divergence of the quantity $z$ and of the quantity $y$ or,
equivalently, the vanishing of the scale factor $a$. As we also noted,
both of the singular and non-singular solutions can traverse the $\phi$
axis and extend at the upper as well as at the lower $z$ quarter-plane. The
value $z=0$, for $k=+1$, is an acceptable value for both families since they
obey the same equation (\ref{trans}). The dependence of the scale factor of
the universe is displayed in Figure 2 where the avoidance of the initial as
well as of the final singularity of the closed universe is obvious. Both of
Figures 1 and 2 have been drawn for the choice $s=+1$. According to our
argument of section 2, the solutions, singular or not, of the equations
(\ref{basicz})-(\ref{basicy}) are invariant under the simultaneous change
of $z$ and $s$. This means that if we choose $s=-1$, the Figures 1 and 2
will remain unchanged apart from the sign of $z$ in Figure 1.

The asymptotic form of the non-singular solutions, for the case $k=+1$, for
early and late times can be found by making use of the ansatz
\be
\phi=\phi_0+\phi_1\,t^\beta \qquad , \qquad
\omega=\omega_0+\omega_1\,t^a \,\,.
\label{ansatz1}
\ee
Substituting the above expressions in the system of equations
(\ref{eq1})-(\ref{eq3}), we find that the only acceptable non-singular
solution has the form
\be
\phi=\phi_0 +\frac{2t}{\sqrt{\delta}}
\qquad , \qquad \omega=\frac{1}{2}\,ln\biggl(\frac{3\delta}{2}\biggl)+
\frac{\omega_1}{\sqrt{|t|}} \,\,,
\label{ansatz2}
\ee
where $\phi_0$ and $\omega_1$ are arbitrary constants.
According to the above solution, the asymptotic regions $\phi \rightarrow
\pm\infty$ correspond exactly to $t \rightarrow \pm\infty$ modulo a
constant coefficient. Then, as it is obvious
from Fig.~2 as well, the non-singular solutions are characteri\-zed in both
limits by the same constant $\omega$ parametrized in terms of the parameter
$\delta$, or equivalently by the same constant scale factor $a$,  which
means that our solutions interpolate between the same static Einstein universe
without undergoing a cosmological singularity, either initial of final.

A family of non-singular solutions, for the choice $k=-1$, is depicted in
Figures 3 and 4. As we note in Figure 3, the singularity-free solutions
cover the whole positive $z$ half-plane without being able to cross the
$\phi$-axis since this would lead to an imaginary value of $x$ according
to eq.~(\ref{trans}). The singular solutions, although not shown in
Figure 3, cover the lower-right-hand-side quarter-plane of the graph, for
the choice $\delta=0.5>0$, and they are completely separated from the
non-singular ones. In Figure 4, the dependence of the scale factor $a$ on
the scalar field is displayed with the absence of the initial singularity
of the open universe being obvious. The Figures 3 and 4 have been drawn
for the choice $s=-1$. As in the previous case, the alternative choice
$s=+1$ leads to the same graphs apart from the sign of $z$ due to the
corresponding invariance of the equations (\ref{basicz})-(\ref{basicy}).
According to Fig.~4, the scale factor adopts a non-vanishing, constant
value in the limit $\phi \rightarrow \infty$ while it increases rapidly
as $\phi \rightarrow -\infty$. Given that the dependence of the scalar field
$\phi$ on time $t$ is a smooth function leading to the asymptotic behaviour
$\phi \rightarrow \pm\infty$ when $t \rightarrow \mp\infty$, we may
conjecture that the non-singular open universe interpolates between a static
Einstein universe at early times and an expanding universe at late times.

In order to complete the picture, we display the graph of the gravitational
scalar curvature $R=g^{\mu\nu}\,R_{\mu\nu}=6\,\ddot{\omega}
+12\,{\dot{\omega}}^2+6k e^{-2\omega}$ versus the scalar
field $\phi$, for the cases of the closed and open universe, in Figs.~5 and
6, respectively. The absence of divergences, which implies the absence of
cosmological singularities, is obvious in both figures. For the case of the
closed universe, $k=+1$, the scalar curvature interpolates between two
constant values, as $\phi \rightarrow \pm\infty$, while in the case of
the open universe, $k=-1$, the scalar curvature adopts a constant value,
in the limit $\phi \rightarrow \infty$, reaches a maximum value near
$\phi=0$ and vanishes quickly as we approach the asymptotic region $\phi
\rightarrow -\infty$.

\vspace{8mm}
\section{Conclusions}

In this article, we have studied a general field theory that describes
the coupling of a scalar field to higher-curvature gravity through the
quadratic Gauss-Bonnet term. As we mentioned in the Introduction, the presence
of this term to the action functional of the Superstring Effective Theory
has led to the existence of new black hole solutions and non-singular
cosmological solutions. For reasons that will be clarified shortly, we
have chosen to study a slightly different quadratic gravitational theory
where the coupling function between the scalar field and the Gauss-Bonnet
term has the polynomial form $\xi(\phi)=\phi^n$, with $n$ being a positive 
integer.

The classical, scalar and gravitational, equations of motion were solved
initially in the absence of the Gauss-Bonnet term. Assuming a constant or
a time-evolving scalar field, we determined the cosmological solutions
that correspond to an empty universe or to a universe that is dominated by
the energy-momentum tensor of a scalar field, respectively. In both cases,
the cosmological solutions were found to be characterized by, at least,
one true cosmological singularity which could not be avoided if a realistic
solution was demanded.

In the presence of the Gauss-Bonnet term, the energy conditions were
shown to be violated allowing the existence of non-singular cosmological
solutions. For $k=0,+1$ this violation holds only for $\delta>0$ while
for $k=-1$ the conditions are violated for both signs of $\delta$.
However, only for $\delta >0$ non-singular solutions arise in the context
of the theory. Next, the equations of motion were
solved analytically near the region of the cosmological singularity. A
family of singular cosmological solutions was determined and was shown
to correspond to a linearly expanding universe, $a(t) \sim (t-t_s)$, with
a true cosmological singularity at finite time $t_s$. This singularity can
be considered as an initial or a final one and the corresponding universe
can be interpreted as an open or closed. One of the basic conclusions drawn
from this analysis was that these solutions are confined in a part of the
phase space of the theory leaving the rest of the space for the non-singular
ones. This remarkable feature holds only when the coupling function has the
dual, polynomial form $\xi(\phi)=\phi^n$, where n is an even, integer number.

The specific choice $\xi(\phi)=\phi^2$ was studied numerically and the
solution for the parameter $z$ and the scale factor $a$ was determined.
For both values of the parameter $k$, the singular cosmological solutions
were confined in a part of the phase space of the theory, as predicted,
and completely separated from the non-singular ones. For the case $k=+1$,
was even possible to found the asymptotic, analytic form of the non-singular
solutions in the limit $t \rightarrow \pm\infty$. According to this form
and our numerical results, the universe, for the case $k=+1$, interpolates
between the same asymptotic, static Einstein universe with constant scale
factor passing through an intermediate phase of expansion-contraction.
By making use of the asymptotic form of the non-singular solutions
(\ref{ansatz2}) for early and late times, we may easily find that the 
scale factor asymptotically adopts the value~: $a(t)\equiv e^\omega \simeq
\sqrt{\frac{\textstyle 3\delta}{\textstyle 2}}$. This means that the
coupling parameter $\delta$ plays the role of the asymptotic value of the
scale factor of the universe which explains the appearance of non-singular
cosmological solutions only for positive values of $\delta$. According to
the above, in the limit $\delta \rightarrow 0$, that is when we eliminate
the Gauss-Bonnet term from the action functional of the theory, the asymptotic 
value of the scale factor goes to zero and we recover the singular
cosmological solutions of section 2.
For the case $k=-1$, we may conjecture that the universe interpolates between
a static Einstein universe at early times and an expanding universe at late
times. For both values of $k$, the absence of cosmological singularities, a feature that owes its existence to the presence of the higher-derivative Gauss-Bonnet term, is obvious.

As is well known, similar results arise in the context of the Superstring
Effective Theory where the coupling function between the scalar field and
the Gauss-Bonnet term has the form
\begin{equation}
\xi(\phi)=ln [2\,e^\phi\,\eta^4(ie^\phi)]\,\,,
\end{equation}
where $\eta$ is the Dedekind function. However, the above form and the
polynomial form that we have chosen during this article share a number
of important characteristics. More analy\-tically~: they are both invariant
under the change $\phi \rightarrow -\phi$, they both have a global minimum
at $\phi=0$ and they both take on an infinite value as $\phi \rightarrow
\pm \infty$. As a result, we may state the following ``theorem"~: {\it
Any theory of the form (\ref{action}) that describes the coupling of a
scalar field with the  Gauss-Bonnet term through a coupling function
which (i)~is dual, (ii)~has a global minimum and (iii)~asymptotically
tends to infinity may lead to singularity-free cosmological solutions}. Superstring Effective Theory is a characteristic example of such a theory
but it is not the only one. Another choice, as we have demonstrated both analytically and numerically, is the theory (\ref{action}) with 
$\xi(\phi)=\phi^{2n}$ and a number of alternative choices may follow as
long as the corresponding coupling functions satisfy the three
aforementioned criteria. 

An important subject which has not been addressed in this article is the
stability of our singularity-free solutions. Here, we have focused our
attention to the existence itself of these solutions. The question of
their stability under linear or even non-linear perturbations is an
independent subject which still remains open. If our singularity-free
solutions turn out to be stable, that will constitute a great achievement
towards the resolution of the initial singularity problem. If, on the other
hand, these solutions turn out to be unstable, their collapse in the early 
universe may have led to the creation of primordial black holes~\cite{kawai}.
We hope to come back to these open questions in a future publication.   

\vspace*{8mm}

\noindent
{\Large \bf Acknowledgements}\\[1mm]

P. K. would like to acknowledge financial support from the research program
$\Pi$ENE$\Delta$-95 of the Greek Ministry of Science and Technology.

\newpage

\begin{figure}
\centerline{\epsfxsize=5in \epsfbox{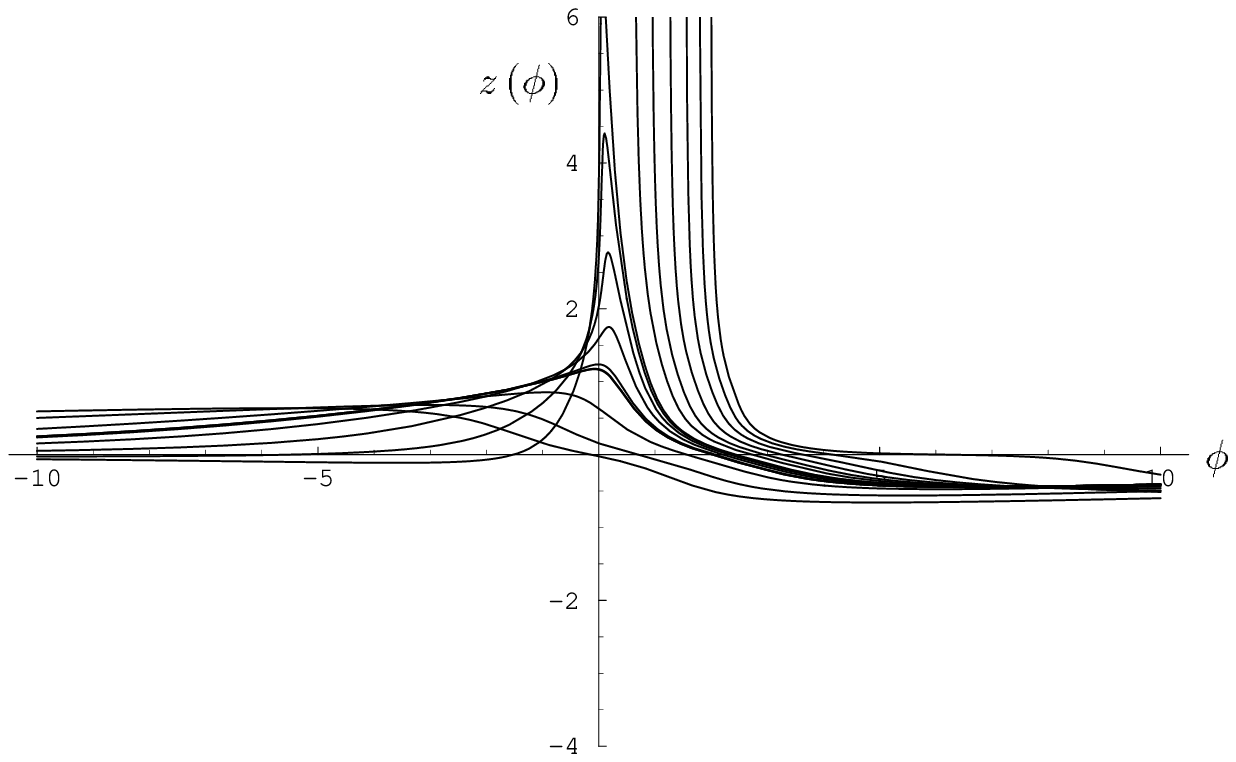}}
\vspace*{2mm}
\caption{The dependence of the quantity $z$ on the scalar field $\phi$ for
a family of singular and non-singular solutions for the case
$k=+1$, $s=+1$ and $\delta=0.5$. }
\end{figure}
\begin{figure}
\centerline{\epsfxsize=5in \epsfbox{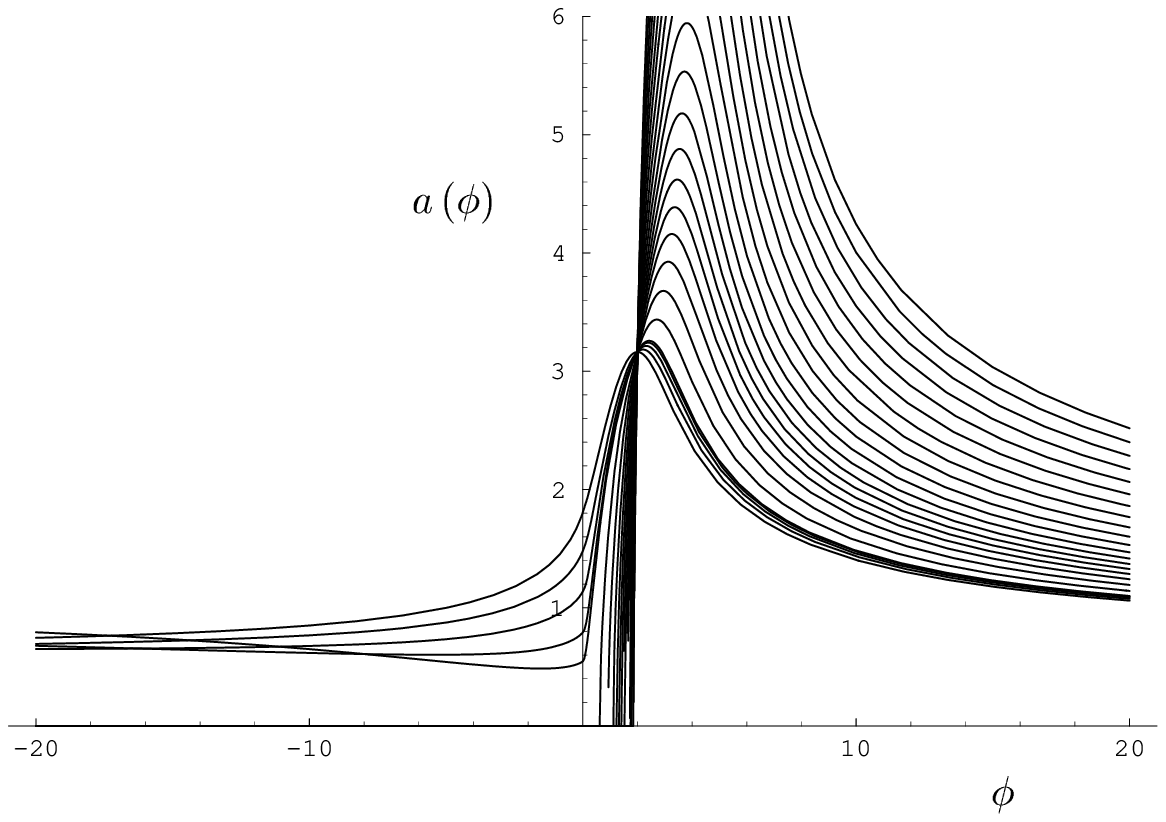}}
\vspace*{2mm}
\caption{The dependence of the scale factor $a$ of the universe on the
scalar field $\phi$ for a family of singular and non-singular solutions
for the case $k=+1$, $s=+1$ and $\delta=0.5$.}
\end{figure}
\begin{figure}
\centerline{\epsfxsize=5in \epsfbox{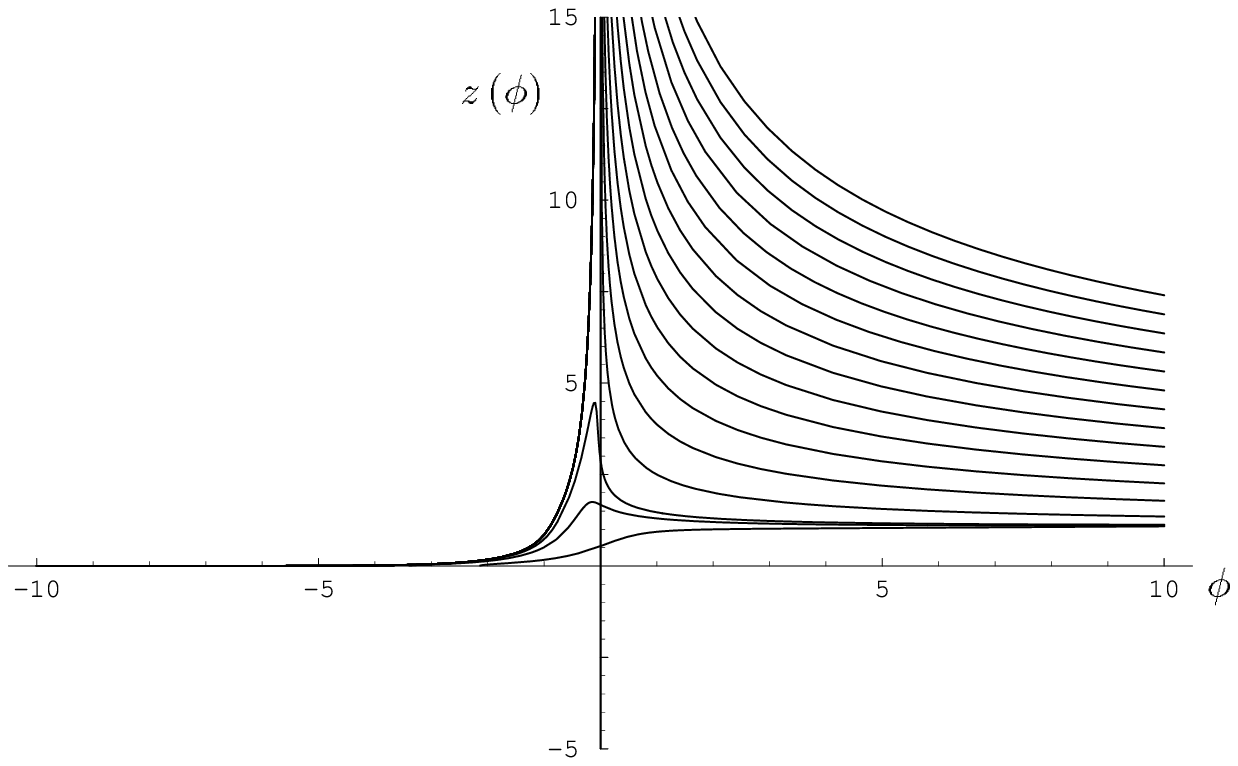}}
\vspace*{2mm}
\caption{The dependence of the quantity $z$ on the scalar field for a
family of non-singular solutions for the case $k=-1$, $s=-1$ and
$\delta=0.5$.}
\end{figure}
\begin{figure}
\centerline{\epsfxsize=5in \epsfbox{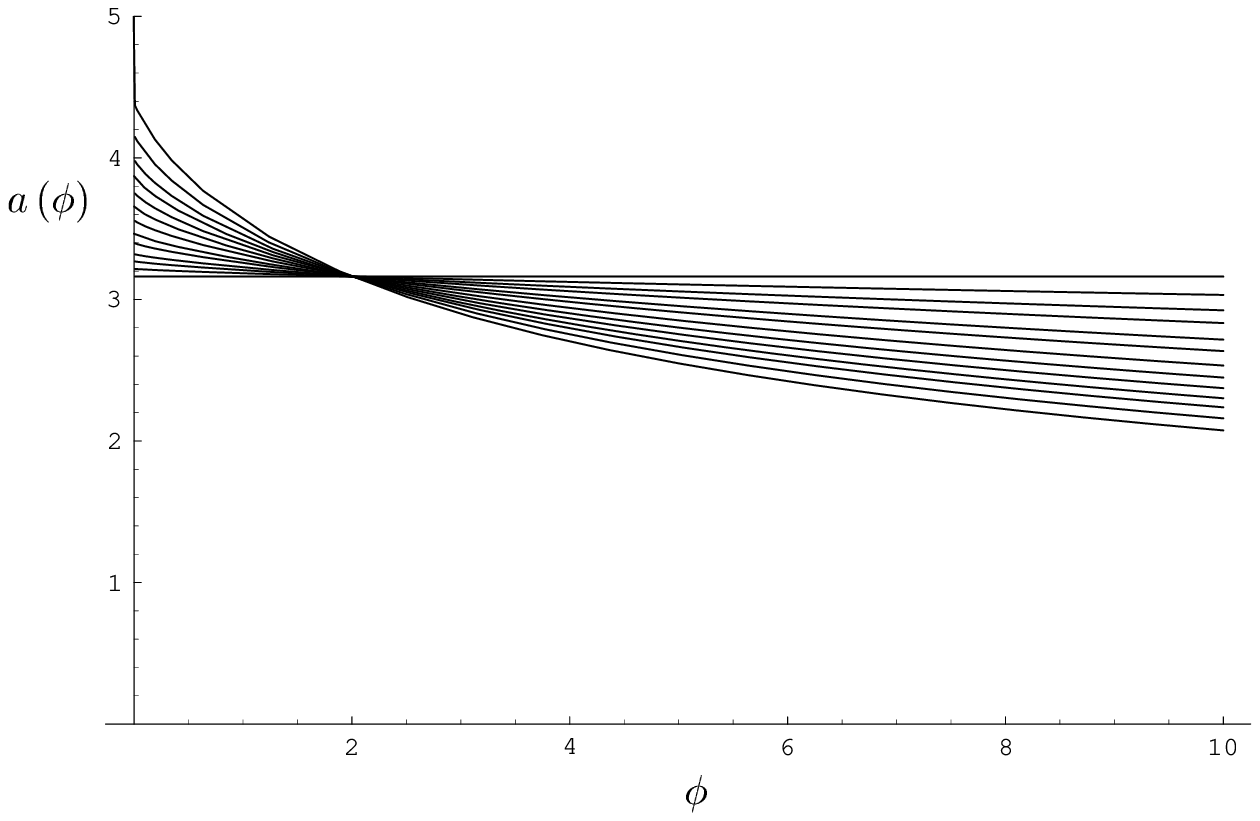}}
\vspace*{2mm}
\caption{The dependence of the scale factor $a$ on the scalar field for
a family of non-singular solutions for the case $k=-1$, $s=-1$ and
$\delta=0.5$.}
\end{figure}
\begin{figure}
\centerline{\epsfxsize=5in \epsfbox{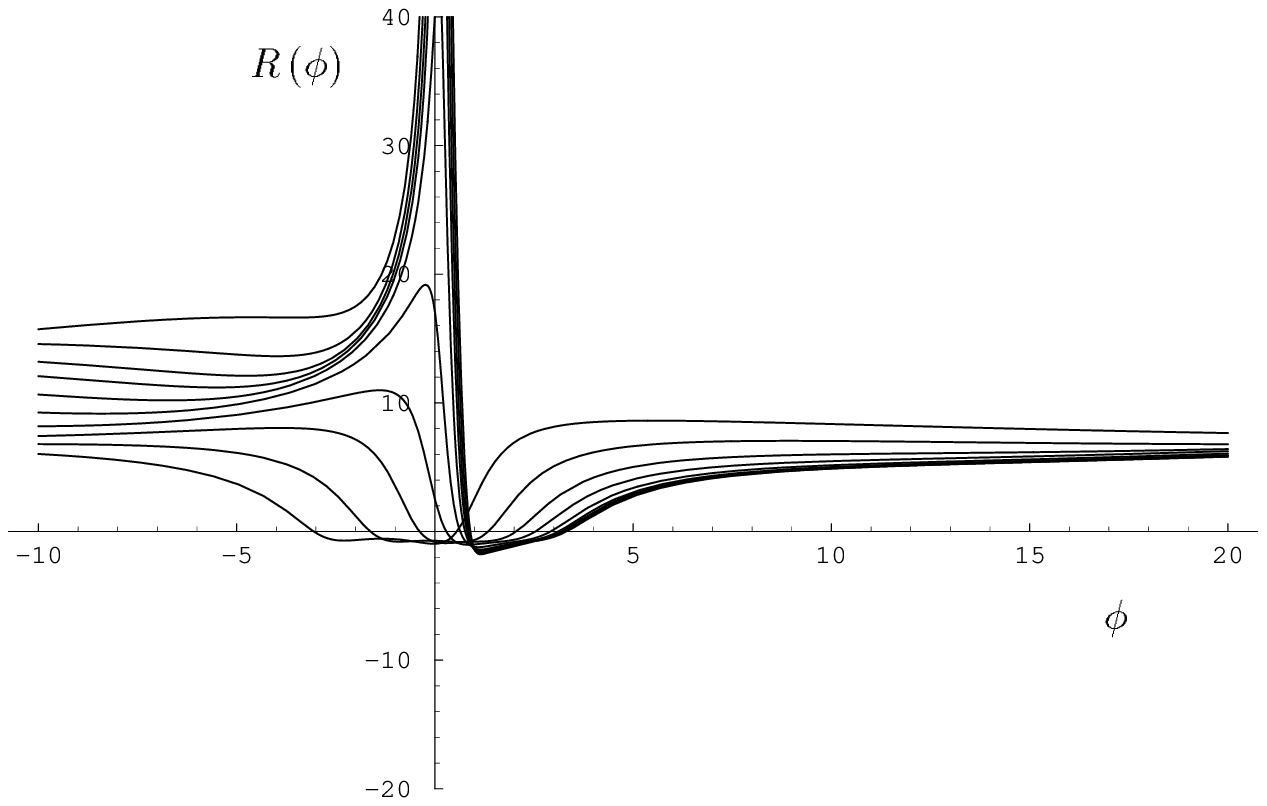}}
\vspace*{2mm}
\caption{The scalar curvature $R$ versus the scalar field $\phi$ for a
family of non-singular solutions for the case $k=+1$, $s=+1$ and
$\delta=0.5$.}
\end{figure}
\begin{figure}
\centerline{\epsfxsize=5in \epsfbox{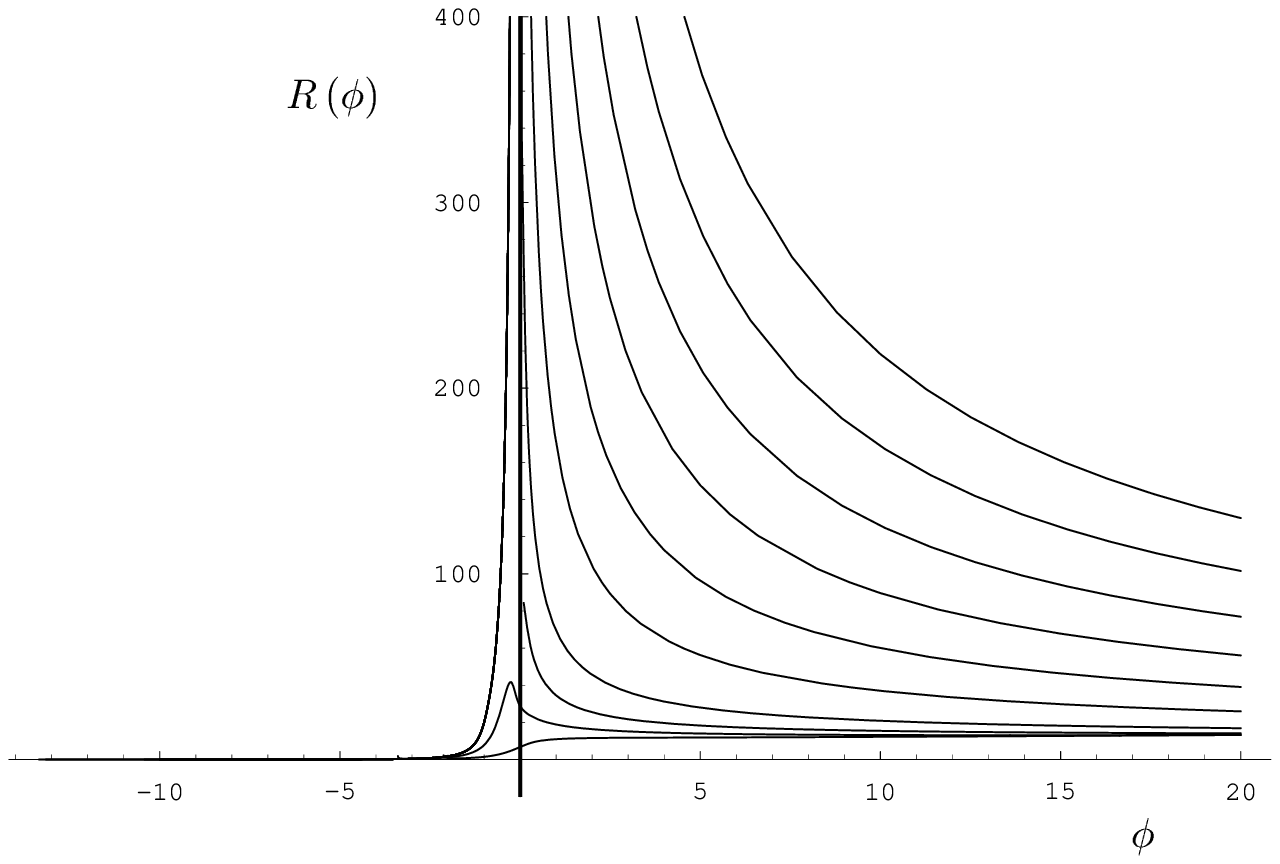}}
\vspace*{2mm}
\caption{The scalar curvature $R$ versus the scalar field $\phi$ for a
family of non-singular solutions for the case $k=-1$, $s=-1$ and
$\delta=0.5$.}
\end{figure}

\end{document}